\documentstyle[prc,aps,epsf,twocolumn]{revtex}

\newcommand{\lag}{\langle}
\newcommand{\rag}{\rangle}

\newcommand{\PRC}[3]{Phys. Rev. C {\bf #1}, #3 (#2)}
\newcommand{\PRL}[3]{Phys. Rev. Lett. {\bf #1}, #3 (#2)}
\newcommand{\NPA}[3]{Nucl. Phys. A {\bf #1}, #3 (#2)}
\newcommand{\PLB}[3]{Phys. Lett. B {\bf #1}, #3 (#2)}
\newcommand{\ZPA}[3]{Z. Phys. A {\bf #1}, #3 (#2)}
\newcommand{\PST}[3]{Physica Scripta T {\bf #1}, #3 (#2)}
\newcommand{\PTP}[3]{Prog. Theor. Phys. {\bf #1}, #3 (#2)}

\newcommand{\ibid}[3]{{\it ibid.} {\bf #1}, #3 (#2)}

\makeatletter
\def\abstract#1{\long\def\@abstract{#1}}
\def\pacs#1{\long\def\@pacs{#1}}
\def\@abstract{}
\def\@pacs{}
\let\@oldmaketitle\@maketitle
\def\@maketitle{%
  \@oldmaketitle
  \begin{quotation}\@abstract\end{quotation}%
  \begin{quotation}\noindent\@pacs\end{quotation}}
\makeatother

\begin{document}

\draft

\title{Relativistic mean field description \\
for the shears band mechanism in $^{84}$Rb}
\author{Hideki Madokoro,$^{1,}$%
\thanks{Electronic address: madokoro@postman.riken.go.jp}
Jie Meng,$^{2,}$%
\thanks{Electronic address: mengj@pku.edu.cn}
Masayuki Matsuzaki,$^{1,3,}$%
\thanks{Electronic address: matsuza@fukuoka-edu.ac.jp}
and Shuhei Yamaji$^{1,}$%
\thanks{Electronic address: yamajis@rikaxp.riken.go.jp}}
\address{$^{1}$RI Beam Factory Project Office, RIKEN, Wako, Saitama 351-0198,
  Japan, \\ $^{2}$Department of Technical Physics, Peking University, Beijing
  100871, China, \\ $^{3}$Department of Physics, Fukuoka University of
  Education, Munakata, Fukuoka 811-4192, Japan}
\date{June 6, 2000}
\abstract{For the first time, the Relativistic Mean Field (RMF) model is
applied to the shears band recently observed in $_{37}^{84}$Rb.
Signals of the appearance of the shears mechanism, such as smooth decreases
of the shears angle and of the $B$(M1)/$B$(E2) ratio with keeping the nearly
constant tilt angle, are well reproduced.  Thus it is shown that the
microscopic RMF model can nicely describe the shears band in this nucleus.}

\maketitle



Recently obtained data of the so-called shears bands in the proton rich
Pb-isotopes\cite{ref:Cl92,ref:Ba92,ref:Ba94,ref:Ne95,ref:Hue97} are well
described within the framework of the Tilted Axis Cranking (TAC)
\cite{ref:KeO81,ref:FrBe92,ref:Fr93,ref:FrMeRe94,ref:FaMe96,ref:Fr00} approach.
In the shears bands, the magnetic dipole vector, which arises from few proton
particles (holes) and few neutron holes (particles) in high-$j$ orbitals,
rotates around the total angular momentum vector.  At the band head, the
proton and neutron angular momenta are almost perpendicular.  This coupling
results in the total angular momentum which is tilted from the principal axes.
With an increase of the rotational frequency, both the proton and neutron
angular momenta align toward the total angular momentum.  Consequently, the
direction of the total angular momentum does not change so much and regular
rotational bands are formed in spite of the fact that the density distribution
of the nucleus is almost spherical or weakly deformed.  These kinds of
rotation are called magnetic rotation\cite{ref:FrMeRe94,ref:Fr97} in order
to distinguish from the usual collective rotation in well-deformed nuclei
(called electric rotation).  Magnetic rotation has also been observed in
other regions such as
$A\sim 110$\cite{ref:Ga97,ref:Cha97,ref:Va98,ref:Je98,ref:Cl99,ref:Kel00}
and 140\cite{ref:Br96} regions.  In a recent experiment, new shears bands in
the Rb-isotopes were discovered\cite{ref:Sc98,ref:Doe99}.  These are the first
experimental data obtained in the $A\sim 80$ mass region.  From the
theoretical side, such shears bands have been well examined by the shell
model\cite{ref:FrReWi96} and a model based on the mean field
approximation\cite{ref:Fr93}.  The shell model approach is especially suitable
for those in the vicinity of doubly closed nuclei.  When we add more and more
particles to the doubly closed core, a transition from complicated
multiplet-like level structures to regular rotational bands in deformed nuclei
is observed.  Shears bands are seen in the middle region, that is, the system
in which only few proton particles/holes and few neutron holes/particles are
involved.  The shell model can describe well both a multiplet-like and a
rotational-like structure, including the shears bands, as well as a transition
from one to another structure.

A weak point of the shell model approach is the effects coming from the
truncation of the model space.  As for the study of shears bands, it is
suggested\cite{ref:FrReWi96} that the polarization of the low-$j$ orbitals
(e.g. $pf$-shells in the Pb-isotopes) would have a `glue' role, which keeps
the outside high-$j$ particles or holes being a blade of shears.  This is
important for the bands to be stabilized.  The truncated shell model may not
be suitable to describe such effects properly because of the limitation of
the model space.  On the other hand, such polarization effects can be easily
included in the mean field models as there is no limitation of the
configuration space.  It is difficult, however, to describe the multiplet-like
level structures and the transition from multiplet-like to rotational-like
structures in the simple mean field approach.  Therefore, complementary
approaches, from both the shell model and the mean field models, are necessary
in order to get a whole description of the property of the shears bands.

Focusing on the mean field approaches to the shears bands, only studies based
on the pairing+QQ Hamiltonian\cite{ref:Fr93,ref:Ba94} have been done up to now.
Those based on more sophisticated models, such as the Skyrme Hartree Fock (SHF)
and the Relativistic Mean Field (RMF) models are still missing and strongly
desired.  In a previous preliminary work\cite{ref:MaMeMaYa00}, we have applied
the RMF model to the tilted axis rotation.  In the present paper, we for the
first time describe the shears bands by the RMF model, which has gratefully
been successful in reproducing many properties of finite nuclei.

In the RMF model\cite{ref:Ri96}, we consider the following Lagrangian, which
contains the nucleon and several kinds of meson fields, such as $\sigma$-,
$\omega$- and $\rho$-mesons, together with the photon fields (denoted by $A$)
mediating the Coulomb interaction:
\begin{displaymath}
  {\cal L}={\cal L}_{N}+{\cal L}_{\sigma}+{\cal L}_{\omega}+{\cal L}_{\rho}
  +{\cal L}_{A}+{\cal L}_{\rm int}+{\cal L}_{\rm NL\sigma},
\end{displaymath}

\noindent
here ${\cal L}_{\rm int}$ is the interaction part between nucleons and mesons.
The non-linear self interactions among the $\sigma$-mesons,
${\cal L}_{\rm NL\sigma}$, are also included.

For applications to rotating nuclei, the Lagrangian of the RMF model must be
generalized into a uniformly rotating frame which rotates with a constant
rotational frequency, $\bbox{\Omega}=(\Omega_{x},\Omega_{y},\Omega_{z})$.
From this generalized Lagrangian, the equations of motion are derived.  This
can be done in the same manner as that in the Principal Axis Cranking (PAC)
case\cite{ref:KoRi89,ref:KaNaMa93,ref:AfKoeRi96,ref:MaMa97,ref:MaMeMaYa00}.
The resulting equations are
\begin{eqnarray*}
  \left\{
    \bbox{\alpha}\cdot(\frac{1}{i}\bbox{\nabla}-g_{\omega}
    \bbox{\omega})+\beta(M-g_{\sigma}
    \sigma)\right. & & \\
    \left. +g_{\omega}\omega^{0}
    -\bbox{\Omega}\cdot(\bbox{L}+\bbox{\Sigma})\raisebox{0pt}[0pt][10pt]{}
    \right\}
  \psi_{i} & = & \epsilon_{i}\psi_{i}, \\
  \left\{
    -\bbox{\nabla}^{2}+m_{\sigma}^{2}-(\bbox{\Omega}\cdot\bbox{L})^{2}
  \right\}\sigma-g_{2}\sigma^{2}+g_{3}\sigma^{3}
  & = & g_{\sigma}\rho_{s}, \\
  \left\{
    -\bbox{\nabla}^{2}+m_{\omega}^{2}-(\bbox{\Omega}\cdot\bbox{L})^{2}
  \right\}
  \omega^{0}  & = & g_{\omega}\rho_{v}, \\
  \left\{
    -\bbox{\nabla}^{2}+m_{\omega}^{2}-(\bbox{\Omega}
    \cdot(\bbox{L}+\bbox{S}))^{2}
  \right\} \bbox{\omega}
  & = & g_{\omega}
  \bbox{j}_{v},
\end{eqnarray*}

\noindent
where the $\rho$-meson and photon fields are omitted for simplicity, although
they are included in the numerical calculation.  It is
known\cite{ref:AfKoeRi96} that the Coriolis terms for the meson and photon
fields (those proportional to $\Omega^{2}$ in the above equations)
give very small contribution and can be completely neglected.  The method
used to solve the coupled equations of motion is, again, the same as that in
the PAC case.  The nucleon and meson fields are expanded in terms of
3-dimensional harmonic-oscillator eigenfunctions.  Note that, contrary to the
PAC case, the signature symmetry is broken in the TAC approach.  The parity
is thus the only symmetry we assume in our code.  The cutoff parameters of
the expansion are taken as $N_{\rm F}=10$ for nucleon fields and
$N_{\rm B}=10$ for meson fields, respectively.  When we increase these cutoff
parameters to $N_{\rm F}=N_{\rm B}=12$, we find the changes of the calculated
values are only $0.1\%$ for the total energies, $2-3\%$ for the total
quadrupole moments and $0.5\%$ for the total angular momenta, respectively.
As for the parameter set, we use that called NL3\cite{ref:LaKoeRi97}.  In the
present code, the pairing correlations are not taken into account.  They
should be included for a precise description of the properties of heavy and
medium-heavy nuclei in the low spin region.  In the relativistic case,
however, the size of the Hamiltonian in the Hartree-Bogoliubov equation
becomes twice as large as that of the non-relativistic case due to the
degrees of freedom of the lower components.  This makes it very time-consuming
to perform a 3-dimensional cranking calculations with pairing.  Because this
is the first RMF work for the shears bands, one of the important purposes of
which is to examine its applicability to the shears bands, we here concentrate
on the calculation without including the pairing interaction.


As the first example of our RMF calculation of the shears bands, we choose
the nucleus $_{37}^{84}$Rb$_{47}$.
In all of our calculations, we assume that the proton configuration is fixed
to be $\pi(pf)^{7}(1g_{9/2})^{2}$ with respect to the $Z=28$ magic number,
that is, a pair of protons align into the $1g_{9/2}$ orbital.  This alignment
is important for the appearance of the shears bands: otherwise no tilted
rotating state is observed.  As for the neutron configuration, we assume
$\nu(1g_{9/2})^{-3}$ with respect to the $N=50$ magic number.  These
assumptions lead to a rotational band with negative parity, which is
consistent with the experimental observation.  In order to find minima, we
increase $\Omega=|\bbox{\Omega}|$ with a step of 0.1 MeV, for each of which
the minimum in the $\theta-\phi$ plane is searched.  Our definition of the
tilt angles is the same as that in Refs. \cite{ref:FaMe97} and
\cite{ref:OhShi97}, that is, $(\theta,\phi)=(90^{\circ},0^{\circ})$ corresponds
to normal collective rotation around the $x$-axis.  This procedure is
continued until $\Omega=0.6$ MeV.  When we come to $\Omega=0.7$ MeV, another
minimum appears at which the shape is almost spherical.  This is consistent
with the experimental observation of the up-bending seen in $J^{(2)}$ and
$B$(M1)/$B(E2)$ (see Figs. \ref{fig:kdmoms} and \ref{fig:bm1be2} below), which
may imply the occurrence of such crossing.  Therefore we add $\Omega=0.65$ MeV
to the mesh points of the frequency in our calculation and the results are
shown only up to this frequency.  Usually we should distinguish the tilt
angles defined for the rotational frequency vector $\bbox{\Omega}$
(denoted by $\theta_{\Omega}$ and $\phi_{\Omega}$) from those defined for
the angular momentum vector $\bbox{J}$ (denoted by $\theta_{J}$
and $\phi_{J}$).  In this study, however, we are interested in only each
minimum, where $\bbox{\Omega}//\bbox{J}$ and therefore ($\theta_{\Omega}$,
$\phi_{\Omega}$) coincides with ($\theta_{J}$,$\phi_{J}$), and so we simply
denote them by $(\theta,\phi)$.  The
deformations $\beta_{2}$ and $\gamma$ are calculated from the quadrupole
moments\cite{ref:AfKoeRi96B}.  We find the deformation slightly decreases from
$\beta_{2}\sim 0.18$ at $\Omega=0.3$ MeV to $\beta_{2}\sim 0.16$ at
$\Omega=0.65$ MeV.  These values are close to the fixed value used in the P+QQ
examination\cite{ref:Sc98}, that is, $\epsilon_{2}=0.14$.  The triaxial
deformation is at most a few degrees and rather small, which can be neglected.
This simplifies our calculation because we can concentrate on the
2-dimensional calculations, where $\phi$ is always set to 0$^{\circ}$.
Besides, we can restrict ourselves to only the range
$\theta=0^{\circ}-90^{\circ}$ thanks to the D$_{2}$
symmetry\cite{ref:FaMe97,ref:OhShi97}.

\begin{figure}[htb]
\begin{center}
  \epsfxsize=7.5cm
  \epsfbox{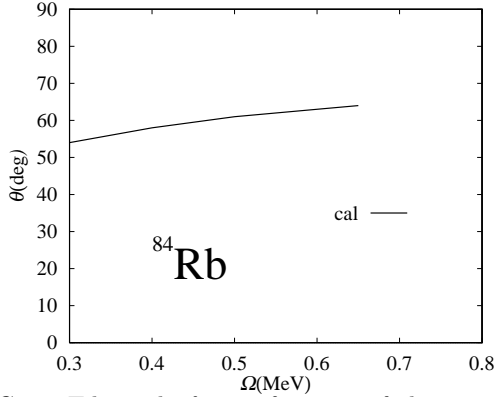}
  \caption{Tilt angle $\theta$ as a function of the rotational frequency.}
  \label{fig:theta}
\end{center}
\end{figure}

Figure \ref{fig:theta} shows how the tilt angle $\theta$ at the minima changes
with respect to the rotational frequency.  At low rotational frequencies, we
find tilted minima appear at $\theta\gtrsim 50^{\circ}$.  With an increase of
$\Omega$, the tilt angle slightly changes but always stays at
$\theta\simeq 55^{\circ}-65^{\circ}$.

In Fig. \ref{fig:kdmoms} the kinematical ($J^{(1)}$) and dynamical ($J^{(2)}$)
moments of inertia are plotted.  The experimental values are calculated from
the transition energies by using the finite difference approximations for
$\Delta I=1$ bands,
\begin{eqnarray*}
  J^{(1)} & = & \frac{I}{E_{\gamma}(I \rightarrow I-1)} \\
    & \mbox{with} & \;\Omega=E_{\gamma}(I \rightarrow I-1), \\
  J^{(2)} & = & \frac{1}{E_{\gamma}(I+1 \rightarrow I)
    -E_{\gamma}(I \rightarrow I-1)} \\
    & \mbox{with} & \;\Omega=\frac{E_{\gamma}(I+1 \rightarrow I)
    +E_{\gamma}(I \rightarrow I-1)}{2}.
\end{eqnarray*}

\noindent
$J^{(2)}$ does not change so much up to $\Omega\sim 0.6$ MeV.
Experimental data show an increase of $J^{(2)}$ above this frequency.  This
might be caused by a crossing as mentioned before.  Although the calculated
$J^{(2)}$ is slightly too large, as a whole both moments of inertia are well
reproduced.  Small discrepancies seen in Fig. \ref{fig:kdmoms} may be cured
by including the pairing correlation.

\begin{figure}[htb]
\begin{center}
  \epsfxsize=7.5cm
  \epsfbox{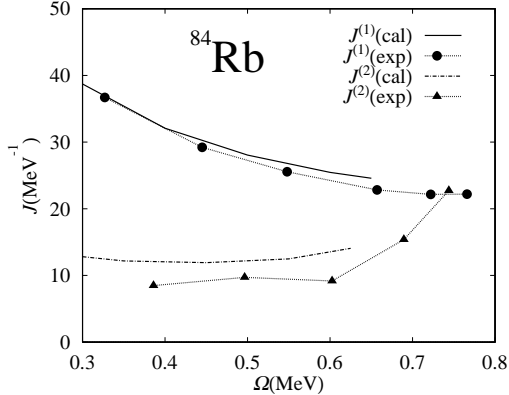}
  \caption{Calculated kinematical ($J^{(1)}$) and dynamical ($J^{(2)}$)
  moments of inertia together with the experimental data taken from Ref.
  \protect\cite{ref:Sc98}.  Solid and dot-dashed lines represent the calculated
  $J^{(1)}$ and $J^{(2)}$, respectively.  Experimental data are shown by
  filled circles ($J^{(1)}$) and triangles ($J^{(2)}$) connected by dotted
  lines.}
  \label{fig:kdmoms}
\end{center}
\end{figure}

\begin{figure}[htb]
\begin{center}
  \epsfxsize=7.5cm
  \epsfbox{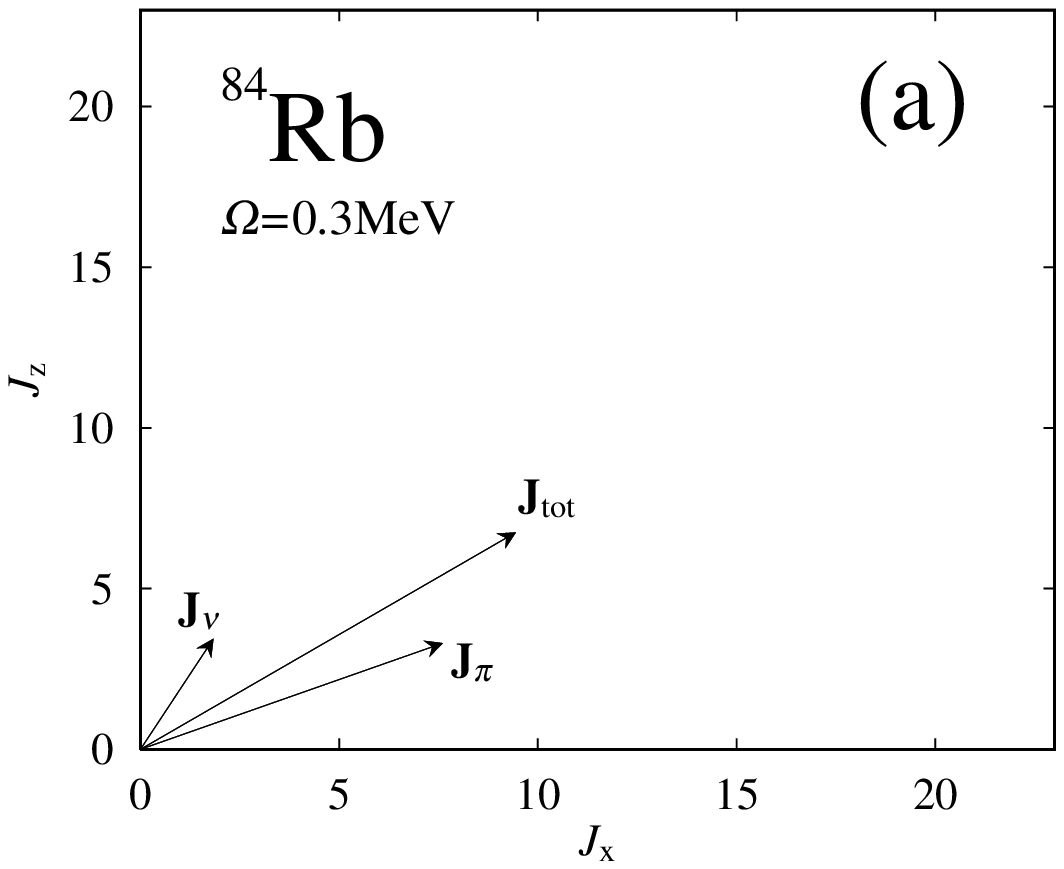}
  \epsfxsize=7.5cm
  \epsfbox{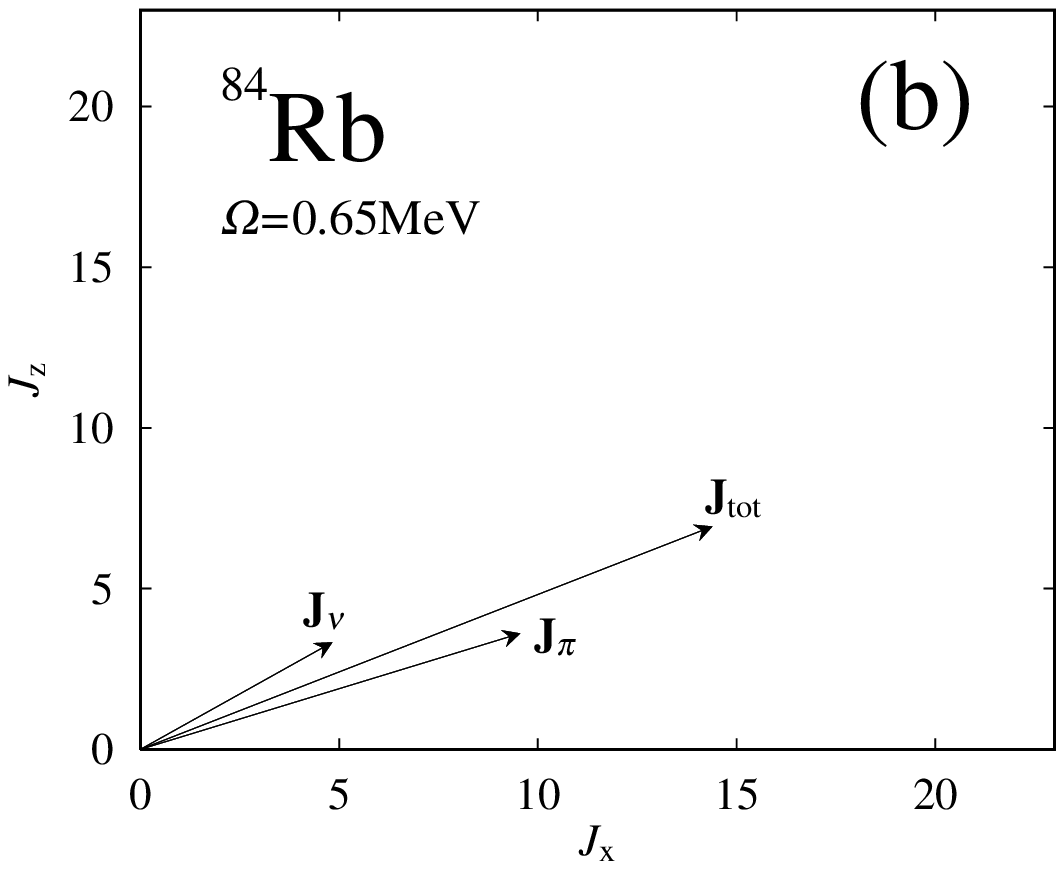}
  \caption{Composition of the total angular momentum at $\Omega=0.3$ MeV (a)
  and 0.65 MeV (b).  {\protect\normalsize $\bbox{J}_{\pi}$} and
  {\protect\normalsize $\bbox{J}_{\nu}$} represent the contributions from
  protons and neutrons, respectively.  {\protect\normalsize
  $\bbox{J}_{\rm tot}$} is the total angular momentum.  The angle between
  {\protect\normalsize $\bbox{J}_{\rm tot}$} and the $J_{z}$-axis coincides
  with the tilt angle $\theta$ (see Fig. \protect\ref{fig:theta}).}
  \label{fig:angvec84Rb}
\end{center}
\end{figure}

Figure \ref{fig:angvec84Rb} shows how the direction of the total angular
momentum vector changes with an increase of the rotational frequency.  Also
shown are the net contributions from protons and neutrons individually.
Because we assume no core in our calculation, $\bbox{J}_{\pi}$ and
$\bbox{J}_{\nu}$ are here defined as the contributions from {\it all}
particles below the Fermi level,
\begin{eqnarray*}
  \bbox{J}_{\pi} & = & \sum_{i=1}^{Z}\lag\bbox{\hat{J}}\rag_{i}, \\
  \bbox{J}_{\nu} & = & \sum_{i=1}^{N}\lag\bbox{\hat{J}}\rag_{i}, \\
  \bbox{J}_{\rm tot} & = & \bbox{J}_{\pi}+\bbox{J}_{\nu}, \\
  |\bbox{J}_{\rm tot}| & \equiv & \sqrt{I(I+1)}.
\end{eqnarray*}

\begin{figure}[htb]
\begin{center}
  \epsfxsize=7.5cm
  \epsfbox{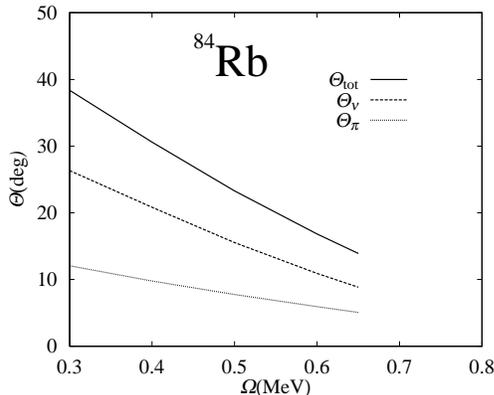}
  \caption{The shears angles $\Theta_{\pi}$, $\Theta_{\nu}$ and
  $\Theta_{\rm tot}$.  $\Theta_{\pi}$ is the angle between {\protect\normalsize
  $\bbox{J}_{\pi}$} and {\protect\normalsize $\bbox{J}_{\rm tot}$},
  $\Theta_{\nu}$ between {\protect\normalsize $\bbox{J}_{\nu}$} and
  {\protect\normalsize $\bbox{J}_{\rm tot}$}, and $\Theta_{\rm tot}$ between
  {\protect\normalsize $\bbox{J}_{\pi}$} and {\protect\normalsize
  $\bbox{J}_{\nu}$}.}
  \label{fig:shearsangle}
\end{center}
\end{figure}

\noindent
We find, however, almost all contributions come from 2 proton particles and
3 neutron holes in the $1g_{9/2}$ orbital.  Several proton particles in the
lower ($pf$) orbitals also give substantial contributions to the proton
angular momentum.  Because of this, $\bbox{J}_{\pi}$ has not only large
$J_{x}$-component but also substantial $J_{z}$-component even at lower
frequencies.  As the frequency increases, we can see that the shears angles
decrease with keeping the direction of the total angular momentum stayed
with nearly constant tilt angle.  This is explicitly shown in Fig.
\ref{fig:shearsangle}.  Here the shears angles $\Theta_{\pi}$, $\Theta_{\nu}$
and $\Theta_{\rm tot}$ are calculated from the following semi-classical
expressions,

\begin{eqnarray*}
  \cos\Theta_{\pi} & = & \frac{\bbox{J}_{\pi}\cdot\bbox{J}_{\rm tot}}
  {|\bbox{J}_{\pi}||\bbox{J}_{\rm tot}|},\;
  \cos\Theta_{\nu}=\frac{\bbox{J}_{\nu}\cdot\bbox{J}_{\rm tot}}
  {|\bbox{J}_{\nu}||\bbox{J}_{\rm tot}|}, \\
  \Theta_{\rm tot} & = & \Theta_{\pi}+\Theta_{\nu}.
\end{eqnarray*}

\noindent
Clearly our result shows almost linear decreases of $\Theta_{\pi}$,
$\Theta_{\nu}$ and $\Theta_{\rm tot}$.  We thus have observed that the shears
mechanism does appear.

Another quantity considered as a signal of the shears mechanism is the ratio
of $B$(M1) to $B$(E2)\cite{ref:Doe87}.  A smooth decrease of the
$B({\rm M1})/B({\rm E2})$ ratio is expected in the shears
bands\cite{ref:Fr93,ref:FrMeRe94}.  These transition probabilities are
calculated from the semi-classical expressions\cite{ref:FaMe97}.  For $B$(M1),
we must calculate the magnetic moments,
\begin{displaymath}
  \bbox{\mu}=g_{l}\sum_{i=1}^{N {\rm or} Z}
    \lag\bbox{\hat{L}}\rag_{i}
    +g_{s}^{\rm (eff)}\sum_{i=1}^{N {\rm or} Z}
    \lag\bbox{\hat{S}}\rag_{i},
\end{displaymath}

\noindent
separately for protons and neutrons.  As for the $g$-factors we use the
standard values\cite{ref:BoMo69}; $g_{l}=1$, $g_{s}=5.58$ for protons and
$g_{l}=0$, $g_{s}=-3.82$ for neutrons, respectively.  Note that
the effective spin $g$-factor ($g_{s}^{\rm (eff)}$) is equal to the free spin
$g$-factor multiplied by 0.7.  We do not introduce the effective charge for
$B$(E2) because our calculation is a fully microscopic one.

\begin{figure}[htb]
\begin{center}
  \epsfxsize=7.5cm
  \epsfbox{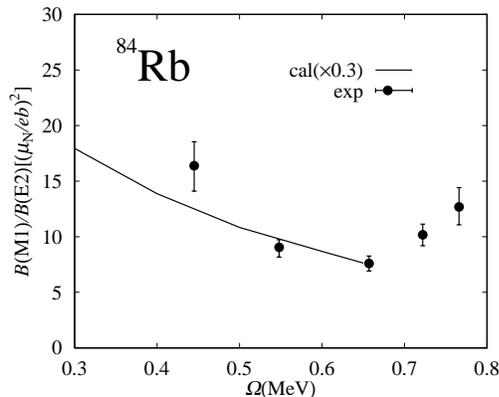}
  \caption{The ratio of $B$(M1) to $B$(E2) as a function of the rotational
  frequency.  The calculated value is multiplied by 0.3.  Experimental data
  are taken from Ref. \protect\cite{ref:Sc98}.}
  \label{fig:bm1be2}
\end{center}
\end{figure}

\noindent
Figure \ref{fig:bm1be2} shows the $B({\rm M1})/B({\rm E2})$ ratio.  The
calculated result is attenuated by a factor of 0.3, because only the $\Omega$
dependence is reliable in the present calculation due to the following reasons:
\begin{enumerate}
\renewcommand{\labelenumi}{(\arabic{enumi})}
\item
  In our results the effect of the pairing interaction is not taken into
  account.  As for $B$(M1), almost all contributions come from the valence
  particles/holes, while in the case of $B$(E2) the contribution from the
  `core' part is rather large.  Because the pairing correlation strongly
  affects the levels near the Fermi level, $B$(M1) is largely affected by
  including the pairing correlation, while $B$(E2) is not.
  In our case of $^{84}$Rb, the tendency that proton particles in the
  low-$k$ levels favor the $x$-axis (rotational-aligned) and neutron holes
  in the high-$k$ levels favor the $z$-axis (deformation-aligned) might be
  smeared out to some extent by including the pairing, because the pairing
  mixes many levels.  This can cause the decrease of the shears angle, and
  then $B$(M1) would be reduced.
  Thus we can expect that including the pairing will
  reduce the $B({\rm M1})/B({\rm E2})$ ratio.
\item
  The particle-vibration coupling calculations based on the RPA show that 
  the in-band $B$(M1)/$B$(E2) values are reduced if the gamma vibration is
  collective enough\cite{ref:MaShiMa88} because it introduces wobbling and
  consequently the overlap between the initial and the final states of the
  transition is reduced.  This would apply also to the present case, as the
  P+QQ result\cite{ref:Sc98} suggests that $^{84}$Rb is a gamma-soft nucleus.
\end{enumerate}

\noindent
The tendency of a smooth decrease of $B({\rm M1})/B({\rm E2})$ up to $\Omega
\sim 0.65$ MeV is well reproduced as can be seen from Fig. \ref{fig:bm1be2}.
Thus our calculation can again show the appearance of the shears bands.


To summarize, we have applied the RMF model to the newly discovered shears
band in $^{84}$Rb.  To this time only studies based on the P+QQ Hamiltonian
had been performed as for the mean field approach to the shears bands.
This is the first examination which is based on a more sophisticated mean field
model.  Our RMF calculation shows decreases of the shears angles and
the $B({\rm M1})/B({\rm E2})$ ratio as the frequency increases, with
keeping nearly constant tilt angle.  Thus we could reproduce the appearance of
the shears mechanism in this nucleus.

Finally, we just give a comment on the case of $^{82}$Rb, in which a shears
band is observed, too. Unfortunately, we could not reproduce the shears band
so definitely for $^{82}$Rb.  A possible reason is as follows: In the
calculation of $^{82}$Rb without pairing, two additional holes must be created
compared with $^{84}$Rb, that is, there exist 5 neutron holes with respect to
the $N=50$ magic number.  It seems that one of these additional two holes
strongly favors $\theta=90^{\circ}$, and therefore works in the direction of
shifting the angular momentum vector toward $\theta=90^{\circ}$.  Pairing
would reduce this effect.

In the $A\sim 80$ region, several high-$K$ bands have been
observed\cite{ref:TaDoe95}.  It remains unsolved, however, whether these are
really shears bands or usual high-$K$ bands in the well deformed nuclei.  In
future, we will examine other isotopes in this region, after including the
pairing correlation, to clarify the question above.

One of us (JM) thanks the hospitality of the Cyclotron Center, RIKEN where
part of this work was done.  His work was also partly supported by the Major
State Basic Research Development Program Under Contract Number G2000077407 and
the National Natural Science Foundation of China (19847002 and 19935030).

\end{document}